Altmetrics
Jason Priem

Chapter from Cronin, B., & Sugimoto, C. R. (2014). Beyond Bibliometrics: Harnessing Multidimensional Indicators of Scholarly Impact (1 edition). Cambridge, Massachusetts: The MIT Press.

Introduction
This chapter discusses altmetrics (short for "alternative metrics"), an approach to uncovering previously-invisible traces of scholarly impact by observing activity in online tools and systems. I argue that citations, while useful, miss many important kinds of impacts, and that the increasing scholarly use of online tools like Mendeley, Twitter, and blogs may allow us to measure these hidden impacts. Next, I define altmetrics and discuss research on altmetric sources—both research mapping the growth of these sources, and scientometric research measuring activity on them. Following a discussion of the potential uses of altmetrics, I consider the limitations of altmetrics and recommend areas ripe for future research.

What are altmetrics?

*The problem: Ideas do not leave good tracks*
It is beyond the scope of this chapter to trace the intellectual and historical lineage of citation mining; interested readers are encouraged to peruse De Bellis' chapter in this volume for an excellent overview. For present purposes, it is sufficient to assert a fact that is common knowledge: ideas, while invisible, are not untraceable. They leave tracks. For 40 years, bibliometricians have diligently hunted, followed, catalogued, and analyzed a particular kind of track: that humble but powerful "pellet of peer recognition" (Merton, 1988, p. 620), the citation.

As they have done so, though, recognition has grown both inside and outside the bibliometrics community that these tracks do not constitute a comprehensive source of data. Many important users of research do not cite; by some estimates, "Only about 15 to 20% of scientists in the United States have authored a refereed article" (King & Tenopir, 2004, para. 13). Many important artifacts are commonly not cited (MacRoberts & MacRoberts, 2010), notably datasets, an increasingly important scientific product. Garfeld himself pointed out that "[c]itation frequency reflects a journal's value and the use made of it, but there are undoubtedly highly useful journals that are not cited frequently" (Garfield, 1972, p. 535). Citations are often used for purposes not accounted for in Merton's normative conception of science, as the social constructivists have argued (Bornmann & Daniel, 2008). Most importantly, citations are product of the slow, rigid *formal* communication system, while scientific ideas themselves are born, nursed, and raised in messy, fast-moving *informal* invisible colleges (de Solla Price & Beaver, 1966). The heart of scientific communication is neither formal publications nor citations but "visits, personal contacts, and letters" (Bernal, 1944, p. 303). Because of this, citations are typically late visitors to the impact "crime scene" (Bellis, 2009, p. 103).

*Solution: Observe new tracks*
The growing pervasiveness of the web is creating an environment in which scholars and other users create new kinds of tracks that evince once-invisible scholarly activities. Inexorably, the daily work of scholars, like other knowledge workers, is moving online. As it does, the background of scholarship—the dog-eared manuscripts and hallway conversations—is pushed out on to the stage. Many online tools are enjoying dramatic growth in usage—growth that seems likely to continue as a "born–digital" generation moves into tenured positions. Social media tools like reference managers, microblogging, and bookmarking services are becoming increasingly important in scholars' workflows, as several recent studies have made clear: Thirteen percent of U.K. academics frequently use Web 2.0 in novel forms of scholarly communications (Procter et al., 2010), and 80% of scholars have social media accounts (Tinti-Kane, Seaman, & Levy, 2010); A U.K. study reports that 10% of doctoral students "use and value" Twitter for research, and that social media tools are affecting the scholarly workflow (Carpenter, Wetheridge, Smith, Goodman, & Struijvé, 2010).

Importantly, these tools are facilitating scholarly practice, but do not seem to be remaking it. Historically, a researcher might read an article, like it, save it in a box or file, discuss it over lunch with her colleagues, cite it in her next paper, and even formally endorse or recommend it at a conference. Today she might download it from an online journal, save it in her reference manager, discuss it with peers on Twitter and blogs, sign it and recommend it on Faculty of 1000.

The difference is that each of these latter activities leaves traces that can be measured.

Of course, an interest in alternative ways to track the flow of ideas is not new. Many alternatives and complements to citations have been proposed and successfully used in the past. Scientometricians have followed and analyzed all sorts of novel impact traces, including acknowledgments (Cronin & Overfelt, 1994), patents (Pavitt, 1985), mentorships (Marchionini, Solomon, Davis, & Russell, 2006), news articles (Lewison, 2002), usage in syllabi (Kousha & Thelwall, 2008), and many others, separately and in various combinations (Martin & Irvine, 1983). Each of these methods has strengths and weaknesses, and all have produced interesting findings. However, they also share a common weakness—they are difficult and time-consuming to gather, especially when compared with citations gathered using commercial indexes like Thomson Reuters' Web of Science (WoS) and Elsevier's Scopus.

Indeed, the idea of measuring impact online is not itself a new one. Practitioners of webometrics have been pursuing digital tracks for some time (Almind & Ingwersen, 1997). However, scalability has been a concern. Although lauded as a solution to the narrowness and slowness of citation metrics (Cronin, Snyder, Rosenbaum, Martinson, & Callahan, 1998), this method has in many ways failed to live up to early expectations (Thelwall, 2010). This is in no small measure due to the difficulty of collecting data—a process commercial search engines have seen little reason to ease, given their heavy investment in proprietary indexes, as well as the relatively tiny number of users interested in these kinds of searches.

The latest sources of online impact differ from earlier citation alternatives, both off- and on-line, in a key respect: they are far easier to gather automatically, since they are based on *tools* and *environments* with clearly defined borders, data types, and (in many cases) structured Web APIs to allow access (Priem & Hemminger, 2010). This allows them to be used at scale, in ways to date uncommon except for citation (and perhaps patent) data. Building on these open APIs, a number of Web-based software tools have emerged to aggregate, process, and present data from multiple APIs. Early tools included ReaderMeter, CitedIn, ScienceCard and *Public Library of Science* Article Level Metrics (*PLoS* ALM); of these, only *PLoS* ALM seems to remain under heavy development at the time of this writing, and is geared more toward use by journals than individual researchers. Recently, the Webometric Analyst package has added Mendeley references to the data it gathers. Two aggregators of particular note for scientometricians are total-impact[1] and Altmetric[2]. The former aggregates a dozen different altmetrics (including Wikipedia citation, Mendeley use, and Tweets) on multiple scholarly products (including slides, datasets, software, and articles); it is open-source and free. The latter is a commercial product that offers particularly good coverage of Twitter, as well as Mendeley, mainstream news outlets, Reddit, and more. These tools are under rapid development.

*Altmetrics*
Altmetrics (jasonpriem, 2010; Priem, Taraborelli, Groth, & Neylon, 2010) is the study and use of scholarly impact measures based on activity in online tools and environments. Unlike the case of "bibliometrics" or "scientometrics," the term has come to be used in describing both the field of study and the metrics themselves (one might advocate the use of a particular "altmetric.") Since altmetrics is concerned with measuring scholarly activity, it is a subset of scientometrics, except when tracking non-science scholarship. Since it is also concerned with measuring activity on the Web, altmetrics is a proper subset of webometrics. It is distinct in that it focuses more narrowly on online *tools and environments*, than on the Web as a whole. Although in theory any Web tool or environment could support altmetrics, in practice, altmetrics researchers have opportunistically sought out rich data pastures, watered by open APIs. They have also focused on tools with large and growing scholarly use. To date, these have included Wikipedia, social reference managers like Mendeley and CiteULike, blogs, Twitter, and several other tools discussed in the next section.

**Altmetrics sources: a summary of research**
Several sources in the mid to late aughts identified the potnetial of altmetrics and called for further research (Jensen, 2007; Neylon & Wu, 2009; Taraborelli, 2008). Since then, research has proceeded apace. However, it is important to note that altmetrics is still in its infancy. Since "[i]t took approximately a generation (20 years) for bibliographic citation anaysis to achieve acceptability as a measure of academic impact" (Vaughan & Shaw, 2005, p.1315), we should expect nothing different from other new metrics. Although altmetrics research is at a very early stage, several methods have already shown themselves useful in its study. Each of these has been employed in the validation of citation metrics, as is shown in the next section.

---

[1] total-impact.org
[2] http://altmetric.com/

*Methods*

Correlation and prediction with established metrics

Garfield (1979) used a correlational and predictive approach to justify citation counts as measures of individual impact, showing how they predicted future Nobel Prize recipients. Narin summarized early efforts to connect citations and esteem measures (Narin, 1976). This has been a common approach in altmetrics research to date, but should be used with caution. We should not expect or even desire perfect correlation between new metrics and traditional metrics (Sugimoto, Russell, Meho, & Marchionini, 2008); part of the value of altmetrics is the ability to measure forms of impact partly or wholly unrelated to what citation captures. Eysenbach (2012), for example, has pursued predictive work with Twitter, while several studies have examined correlation between citation and reference manager inclusion (Haustein & Siebenlist, 2011; Li, Thelwall, & Giustini, 2011), or between multiple altmetrics and citations (Priem, Piwowar, & Hemminger, 2012; Yan & Gerstein, 2011).

Content analysis

There has been a long tradition of context-analytic studies of citation, beginning with Moravcsik and Murugesan's (1975) seminal categorization of citations from 30 high-energy physics articles (see Bornmann and Daniel [2008] for many more examples. Cronin et al. (1998) analyzed web mentions of scholars, and Thelwall and others (Kousha & Thelwall, 2006; Thelwall, 2003) analyzed the context of scholarly hyperlinks. Similar studies have investigated altmetrics, particularly tweets (Letierce, Passant, Breslin, & Decker, 2010; Priem & Costello, 2010; Ross, Terras, Warwick, & Welsh, 2011). Techniques for automatically extracting contextual information can be used to isolate and describe contextual information in tweets (Stankovic, Rowe, & Laublet, 2010).

Creator feedback

Creator feedback studies are also known as "citer motivation" or "citer behavior studies" (Borgman & Furner, 2002). Here, researchers use interviews or surveys to investigate authors' reasons for creating certain types of records. Bornmann and Daniel (2008) review many creator feedback studies of traditional citation beginning with Brooks' (1986) interview-based and Vinkler's (1987) survey-based investigations. More recently, Cronin and Overfelt (1994) used surveys to establish authors' motivations for creating acknowledgments, and Priem and Costello (2010) used interviews to investigate scholars' motivations for citing on Twitter.

Prevalence studies
Although citation practices vary between disciplines, the practice is ubiquitous. The distribution of altmetrics, on the other hand, depends on the uptake of scholarly tools being examined. Consequently, many studies useful to altmetrics have focused less on metrics and more on simply describing the scholarly use of a given environment. Some of these studies are discussed below.

**Notable altmetrics findings by source**
An important property of altmetrics is the ability to track impact on broad or general audiences, as well as on scholars. This is reflected in Table 1, which separates impact types by audience. This section reviews findings relevant to altmetrics, attempting to discuss each cell in Table 1 in turn. I highlight findings that relate specifically to measuring scholarly impact, but also selected studies indicating the prevalence and uptake of the different communication channels and media. Most studies discussed in this section focus on one source; two notable ones that discuss multiple sources use sets of articles from the open-access publisher *Public Library of Science* (*PLoS*) (Priem et al., 2012; Yan & Gerstein, 2011). This is thanks to *PLoS'* Article-Level Metrics (ALM) API, which presents lists of altmetrics data by article, making this an useful test set for early altmetrics investigations. I begin by discussing indicators of impact on the general public. These impacts are vital and not just for research with an immediate public benefit (e.g., medical research). Since the public indirectly funds most research, public awareness and outreach are important, especially in lean economic times.
[insert table 1 here]

*Public*
Mainstream media
Lewison (2002) tracked mass media mentions of scholarly articles, finding little relation between journal citation rates and citedness by major news outlets, suggesting this is a legitimately distinct form of impact. However, other work shows that these mentions seem to affect citation rates, suggesting that the two are not completely unconnected (Kiernan, 2003). Since then, many major news outlets like *The Guardian* and *The New York Times* have launched open APIs that make searching their text much easier, or at least have RSS feeds that can be automatically crawled. Altmetric.com provides a convenient interface for many of these; this should encourage more work tracking this form of impact.

Wikipedia
For much of the world, especially students (Head & Eisenberg, 2010; Schweitzer, 2008), Wikipedia is the first stop for information. Influencing Wikipedia, therefore, means influencing the world in a profound way. Citation on Wikipedia, could be considered a public parallel to scholarly citation (a discussion of the latter is out of scope for this chapter, but it is included in Table 1 for as a reference point). Nielsen (2007) has shown that citations in Wikipedia correlatse well with data from the Journal Citation Report, establishing a relationship between impact on Wikipedia and in more traditional contexts. Priem, Piwowar, and Hemminger (2012) report that around 5% of *PLoS* articles are cited in Wikipedia, and report correlations between .1 and .4 between normalized Wikipedia citations and traditional citations, depending on the journal. Wikimedia Research's Cite-o-Meter tool shows an up-to-date league table of which academic publishers are most cited on Wikipedia.

Conversation (Twitter)
Twitter presents a rich source of data, but a difficult one for investigators of scholarly impact. Unlike many tools discussed in this chapter, Twitter does not make apparent the audience interacting with a research product. Twitter is heavily used by some scholars, as we shall see later, but non-scholars make up the bulk of Twitter users. Given this, we assume tweets come from lay readers unless given evidence to the contrary (such contrary evidence could come from biographical information, follower lists, or contents of tweets; Altmetric.com uses an algorithm to decide). This assumption is supported by the near-zero correlation between tweets and citations found in the *PLoS* dataset (Priem et al., 2012), suggesting that the high visibility of *PLoS* attracts many non-scholarly readers. On the other hand, Eysenbach's (2012) study of *Journal of Medical Internet Research* (JMIR) papers found a distinct link between tweeting and citation: articles in the top quartile of tweets were 11 times more likely to be in the top quartile of citations two years later. This may be because *JMIR* tweeters are more likely to be the same scholars who might cite the work. More research is needed that separates the identity of Twitter users.

Conversation (social news)
Social news includes sites like Reddit, Digg, and Slashdot. Lerman and Galstyan (2008) demonstrated that early "Diggs" predict later importance of news stories, which encourages inquiry into similar predictive validity for scholarly publications. However, while Norman suggests using a "Slashdot index" to measure scholarly impact (Cheverie, Boettcher, & Buschman, 2009), no research has yet emerged tracking scholarly articles' mentions on recommendation sites like these. Reddit is a particularly interesting case, since users can create focused "subreddits." Although these can appear, mutate, and disappear rapidly, at the time of writing, there were subreddits built around requesting paywalled scholarly papers, discussing the latest science research (reporting nearly 2 million readers), and others; subreddits are an interesting subject for future altmetrics research.

Conversation (bookmarking)
The small body of research into social bookmarking of scholarly resources has focused on Delicious[3] (Lund, Hammond, Flack, Hannay, & NeoReality, 2005). Ding, et al. (2009) found that "[s]cientific domains, such as bioinformatics, biology, and ecology are also among the most frequently occurring tags" (p. 2398), suggesting at least some scholars use the service. Priem, Piwowar, and Hemminger (2012) report that about 10% of *PLoS* articles are bookmarked in Delicious.

html views
With the growth of the open access (OA) movement, it has become possible for us to talk meaningfully about numbers of non-scholarly readers. HTML downloads, when reported separately from PDF downloads, may be useful for identifying non-scholarly readers. Priem, Piwowar, and Hemminger (2012) observe that HTML views tend to cluster with other metrics of public impact. This is probably because the general public is more likely to read only the abstract, or to skim the text of the article quickly, while academics are more likely to download and print the paper. Further research investigating the ratio between HTML views and PDF downloads could uncover interesting findings about how the public interacts with the open access (OA) research literature.

---

[3] http://delicious.com

*Scholars*

In addition to tracking scholarly impacts on traditionally invisible audiences, altmetrics hold potential for tracking previously hidden scholarly impacts.

Faculty of 1000

Faculty of 1000 (F1000) is a service publishing reviews of important articles, as adjudged by a core "faculty" of selected scholars. Wets, Weedon, and Velterop (2003) argue that F1000 is valuable because it assesses impact at the article level, and adds a human level assessment that statistical indicators lack. Others disagree (*Nature Neuroscience*, 2005), pointing to a very strong correlation (r = 0.93) between F1000 score and Journal Impact Factor. This said, the service has clearly demonstrated some value, as over two thirds of the world's top research institutions pay the annual subscription fee to use F1000 (Wets et al., 2003). Moreover, F1000 has been to shown to spot valuable articles which "sole reliance on bibliometric indicators would have led [researchers] to miss" (Allen, Jones, Dolby, Lynn, & Walport, 2009, p. 1). In the *PLoS* dataset, F1000 recommendations were not closely associated with citation or other altmetrics counts, and formed their own factor in factor analysis, suggesting they track a relatively distinct sort of impact.

Conversation (scholarly blogging)

In this context, "scholarly blogging" is distinguished from its popular counterpart by the expertise and qualifications of the blogger. While a useful distinction, this is inevitably an imprecise one. One approach has been to limit the investigation to science-only aggregators like ResearchBlogging (Groth & Gurney, 2010; Shema & Bar-Ilan, 2011). Academic blogging has grown steadily in visibility; academics have blogged their dissertations (Efimova, 2009), and the ranks of academic bloggers contain several Fields Medalists, Nobel laureates, and other eminent scholars (Nielsen, 2009). Economist and Nobel laureate Paul Krugman (Krugman, 2012), himself a blogger, argues that blogs are replacing the working-paper culture that has in turn already replaced economics journals as distribution tools.

Given its importance, there have been surprisingly few altmetrics studies of scholarly blogging. Extant research, however, has shown that blogging shares many of the characteristics of more formal communication, including a long-tail distribution of cited articles (Groth & Gurney, 2010; Shema & Bar-Ilan, 2011). Although science bloggers can write anonymously, most blog under their real names (Shema & Bar-Ilan, 2011).

Conversation (Twitter)
Scholars on Twitter use the service to support different activities, including teaching (Dunlap & Lowenthal, 2009; Junco, Heiberger, & Loken, 2011), participating in conferences (Junco et al., 2011; Letierce et al., 2010; Ross et al., 2011), citing scholarly articles (Priem & Costello, 2010; Weller, Dröge, & Puschmann, 2011), and engaging in informal communication (Ross et al., 2011; Zhao & Rosson, 2009). Citations from Twitter are a particularly interesting data source, since they capture the sort of informal discussion that accompanies early important work. There is, encouragingly, evidence that Tweeting scholars take citations from Twitter seriously, both in creating and reading them (Priem & Costello, 2010). The number of scholars on Twitter is growing steadily, as shown in Figure 1. The same study found that, in a sample of around 10,000 Ph.D. students and faculty members at five representative universities, one 1 in 40 scholars had an active Twitter account. Although some have suggested that Twitter is only used by younger scholars, rank was not found to significantly associate with Twitter use, and in fact faculty members' tweets were twice as likely to discuss their and others' scholarly work.

Conversation (article commenting)
Following the lead of blogs and other social media platforms, many journals have added article-level commenting to their online platforms in the middle of the last decade. In theory, the discussion taking place in these threads is another valuable lens into the early impacts of scientific ideas. In practice, however, many commenting systems are virtual ghost towns. In a sample of top medical journals, fully half had commenting systems laying idle, completely unused by anyone (Schriger, Chehrazi, Merchant, & Altman, 2011). However, commenting was far from universally unsuccessful; several journals had comments on 50-76% of their articles. In a sample from the *British Medical Journal*, articles had, on average, nearly five comments each (Gotzsche, Delamothe, Godlee, & Lundh, 2010).

Additionally, many articles may accumulate comments in other environments; the growing number of external comment sites allows users to post comments on journal articles published elsewhere. These have tended to appear and disappear quickly over the last few years. Neylon (2010) argues that online article commenting is thriving, particularly for controversial papers, but that "...people are much more comfortable commenting in their own spaces" (para. 5), like their blogs and on Twitter.

Reference managers

Reference managers like Mendeley and CiteULike are very useful sources of altmetrics data and are currently among the most studied. Although scholars have used electronic reference managers for some time, this latest generation offers scientometricians the chance to query their datasets, offering a compelling glimpse into scholars' libraries. It is worth summarizing three main points, though. First, the most important social reference managers are CiteULike and Mendeley. Another popular reference manager, Zotero, has received less study (but see Lucas, 2008). Papers and ReadCube are newer, smaller reference managers; Connotea and 2Collab both dealt poorly with spam; the latter has closed, and the former may follow. Second, the usage base of social reference managers—particularly Mendeley—is large and growing rapidly. Mendeley's coverage, in particular, rivals that of commercial databases like Scopus and Web of Science (WoS) (Bar-Ilan et al., 2012; Haustein & Siebenlist, 2011; Li et al., 2011; Priem et al., 2012). Finally, inclusion in reference managers correlates to citation more strongly than most other altmetrics. Working with various datasets, researchers have reported correlations of .46 (Bar-Ilan, 2012), .56 (Li et al., 2011), and .5 (Priem et al., 2012) between inclusion in users' Mendeley libraries, and WoS citations. This closer relationship is likely because of the importance of reference managers in the citation workflow. However, the lack of perfect or even strong correlation suggests that this altmetric, too, captures influence not reflected in the citation record. There has been particular interest in using social bookmarking for recommendations (Bogers & van den Bosch, 2008; Jiang, He, & Ni, 2011).

pdf downloads

As discussed earlier, most research on downloads today does not distinguish between HTML views in PDF downloads. However there is a substantial and growing body of research investigating article downloads, and their relation to later citation. Several researchers have found that downloads predict or correlate with later citation (Perneger, 2004; Brody et al., 2006). The MESUR project is the largest of these studies to date, and used linked usage events to create a novel map of the connections between disciplines, as well as analyses of potential metrics using download and citation data in novel ways (Bollen, et al., 2009). Shuai, Pepe, and Bollen (2012) show that downloads and Twitter citations interact, with Twitter likely driving traffic to new papers, and also reflecting reader interest.

**Uses, limitations and future research**

*Uses*
Several uses of altmetrics have been proposed, which aim to capitalize on their speed, breadth, and diversity, including use in evaluation, analysis, and prediction.

Evaluation
The breadth of altmetrics could support more holistic evaluation efforts; a range of altmetrics may help solve the reliability problems of individual measures by triangulating scores from easily-accessible "converging partial indicators" (Martin & Irvine, 1983, p. 1). Altmetrics could also support the evaluation of increasingly important, non-traditional scholarly products like datasets and software, which are currently underrepresented in the citation record (Howison & Herbsleb, 2011; Sieber & Trumbo, 1995). Research that impacts wider audiences could also be better rewarded; Neylon (2012) relates a compelling example of how tweets reveal clinical use of a research paper—use that would otherwise go undiscovered and unrewarded. The speed of altmetrics could also be useful in evaluation, particularly for younger scholars whose research has not yet accumulated many citations. Most importantly, altmetrics could help open a window on scholars' "scientific 'street cred'" (Cronin, 2001, p. 6), helping reward researchers whose subtle influences—in conversations, teaching, methods expertise, and so on—influence their colleagues without perturbing the citation record. Of course, potential evaluators must be strongly cautioned that while uncritical application of any metric is dangerous, this is doubly so with altmetrics, whose research base is not yet adequate to support high-stakes decisions.

Analysis

Altmetrics could greatly benefit students of science, supplying a set of instruments possessing unprecedented scope and granularity. Some have suggested combining multiple metrics using regression analysis (Harnad, 2009) or combining metrics using Principle Component Analysis (Bollen, Van de Sompel, Hagberg, & Chute, 2009) or time-series analysis (Kurtz et al., 2005) to determine "underlying" forms of impact—the impact equivalent of Spearman's *g*. However, in many ways this approach moves in the wrong direction. Impact, as we are reminded elsewhere in this volume, invariably begs the question, "impact on *what*?" As long as there are multiple answers to these questions, reductive approaches will founder. Instead, it may be more useful to think not in terms of underlying impact, but, rather, of different kinds or flavors of impact, based on different audiences, products, and goals (Piwowar, 2012). Priem, Piwowar, and Hemminger (2012) found evidence of five such impact types from cluster analysis of 25k *PLoS* articles (Figure 2), including "popular hits" (cluster C in the figure), and heavily referenced but relatively uncited papers (Cluster B).
[insert figure 2 here]

The temporal resolution of altmetrics has potential to help scientometricians better understand how knowledge moves through the varyingly-dense media of different communication systems. Fiureg 3 gives an example for a single article. It shows that Delicious, CiteULike, and *PLoS* comments are active immediately after the article's publication. There follows a second wave of activity four or so months later, possibly initiated by readers alerted to the article by its first external citation.
[insert figure 3 here]

Prediction
Altmetrics, in most cases, are fast. They often begin accumulating a few days after publication (Kurtz et al., 2005; Priem, Costello, & Dzuba, 2011) and hold potential as leading indicators of other types of use supporting an "embryology of learned inquiry" (Harnad & Carr, 2000, p. 629). Data like these shown in Figure 3 could be aggregated across thousands or millions of articles to produce powerful predictive models.

Forecasting based on social media has already proven surprisingly effective in diverse areas including predicting stock prices, election results, and movie box-office returns (Asur & Huberman, 2010; Bollen, Mao, & Zeng, 2011; Tumasjan, Sprenger, Sandner, & Welpe, 2010). Altmetrics could support the extension of these same techniques to academic trends and research fronts. Wang, Wang, and Xu (2012) show the potential of this approach, identifying emerging research areas using download counts and keyword data from the journal *Scientometrics*. Here the speed of altmetrics compared to traditional citation measures is of tremendous value. Trend detection using traditional citations is rather like predicting the weather after the first few raindrops; altmetrics could be a radar-sized leap forward in our predictive power.

Recommendation

Although article recommendation systems based on citations have been proposed before (McNee, 2006), they have not seen wide usage, largely because of the lag inherent in citation tracking. By the time the citation record has caught up, researchers' interests have often evolved. The speed of altmetrics could be valuable in this area. One could imagine receiving daily, emailed recommendations based on the aggregated judgments of one's trusted peers, helping to repair the "filter failure" (Shirky, 2008) plaguing the modern research reader. A system like, this, if widely used, would supplement the peer review system, harnessing reader-centered teams of reviewers by tracking their interactions with recently-published literature (Neylon & Wu, 2009). Such a "soft peer review" (Taraborelli, 2008) system could aggregate and compile readership metrics and user-generated comments and reviews; this information could help readers judge the article's importance and relevance. While this is unlikely to replace traditional peer review, it could support faster publishing "post-publication peer-review" journals, or add an additional layer of open quality control to traditional publications.

     A more radical use would be to move away from traditional peer review entirely. Given a sufficiently advanced recommender system, one must wonder, why would we need peer review at all? It could be that the aggregated readership, conversation, and collection of one's trusted peers is enough to separate the gold from the dross. Many have suggested "deconstructed" (Smith, 2003) or "decoupled" journals to take advantage of this. Extending the open-access model, such journals would offer paying authors a shopping cart of copyediting, review, translation, and other services. Upon publication, readers could use their own robust filtering systems, built around the aggregated readership, conversation, and collection judgments of their peers, to uncover the most promising material to read. A decentralized peer review system, by leveraging collective usage

information might filter the literature in a way similar to how Google filters the web based on aggregated link in formation.

*Limitations*
Despite the potential of altmetrics, they have not failed to provoke criticism. The concerns have tended to focus on three areas: a lack of theory, ease of gaming, and bias.

It is certainly true that altmetrics lacks a cohesive body of theory; this lack must be remedied before altmetrics can be widely employed. However, it should also be acknowledged that lack of theory, while important to address, has not kept other new metrics from being useful in practice. One needs look no further than bibliometrics itself for examples of this. As late as 1979, Garfield acknowledged gaps in theoretical understanding of citation, but argued that this should not preclude use of bibliometrics: "we still know very little about how sociological factors affect citation rates...On the other hand, we know that citation rates say something about the contribution made by an individual's work, at least in terms of the utility and interest the rest of the scientific community finds in it" (Garfield, 1979, p. 372). The same can as be said for today's altmetrics. The introduction of new sources for impact metrics such as patents (Pavitt, 1985), acknowledgements (Cronin et al., 1998), mentorships (Marchionini et al., 2006), and scholarly hyperlinks (Ingwersen, 1998) has typically preceded robust theoretical underpinnings; indeed, it is difficult to imagine otherwise. A theory of altmetrics should be a priority for the new field, but not a prerequisite.

A second concern is the ease with which altmetrics counts can be manipulated. Again, this is a legitimate concern—but we should not imagine that extant metrics are free from it, either. Any metric will spawn attempts to exploit it (Espeland & Sauder, 2007). The Journal Impact Factor is a noteworthy example of a heavily gamed metric. The increasing importance of this measure has spawned a whole range of tips and tricks for artificially boosting citation scores (Falagas & Alexiou, 2008), including, recently, the formation of "citation cartels" (Davis, 2012; Franck, 1999) in which journals collude to cite one another. In an extreme example, a single scholar's questionable editorial practices were enough to catapult the University of Alexandria—"not even the best university in Alexandria" into the *Times Higher Education* top 200 rankings (Guttenplan, 2010, para. 6). For every extreme case like this, there are likely many more going undetected.

This said, it is certainly true that the relative ease of creating and using social media profiles seems to make generating false data easier for malfeasants. Indeed, artificial inflation of social media metrics is already a well-established practice outside academia. However, successful businesses and tools have evolved immune systems to combat this kind of gaming, in the form of anti-spam and anti-gaming measures. Perhaps the highest-profile of these is Google; with millions of dollars in traffic at stake, advertisers have assaulted Google search results with "black–hat" search engine optimization (Malaga, 2010). While these have not been entirely unsuccessful, they have not significantly undermined users' trust in the value of Google results. Google uses a variety of constantly-evolving algorithms to differentiate spam sites from legitimate ones; similar statistical techniques, or "algorithmic forensics", can help control social media gaming, as well. For instance, the automated WikiScanner tool[4] exposed and helped correct corporate tampering with Wikipedia articles (Borland, 2007). For Twitter users, Twitteraudit scans lists of followers, using tuned algorithms to spot bots. Moving to academia, the SSRN preprint server measures and reports download statistics for articles posted there. These statistics have become important for evaluation purposes in several disciplines, and have consequently attracted gaming (Edelman & Larkin, 2009). SSRN has used algorithmic forensics to detect fraudulent downloads based on the observed properties of millions of legitimate ones. One particular virtue of an approach examining multiple social media ecosystems is that data from different sources can be cross-calibrated, exposing suspicious patterns invisible in a single source; *PLoS* "DataTrust" uses this approach to spot fraudulent altmetrics counts. While additional work in this area is certainly needed, there is evidence to suggest that social metrics, properly filtered and cautiously interpreted, could be relatively robust despite attempts to game them.

    A third concern with altmetrics is that they will be systematically biased—in particular, toward younger or more fad-embracing researchers or outlets. Will altmetrics create a science dominated by shallow, fame-seeking narcissists, constantly chasing the next trend? While a glib "how would that be different from now?" is needlessly cynical, it holds an element of truth: scientists constantly promote themselves and their ideas across many venues, from conferences to mailing lists to press releases to the popular press. If the new metrics reward those

---

[4] http://wikiscanner.virgil.gr

scientists who make the best use of available technology to provoke conversation amongst their peers and capture the imagination of the public, surely this is a bias we can accept. It is far from clear that efficient and successful use of available communication technologies to forcefully advocate one's ideas is something we should recoil from rewarding. Additionally, there is little evidence confirming naïve assumptions about the demographics of researchers who are adopting new communication technologies. Priem, Costello, and Dzuba (2011), for example, find that doctoral students are no more likely to be active on Twitter than university faculty members. This said, we should take pains in using altmetrics to compare like with like, just as we do with citations; a field like digital humanities, where anecdotal reports suggest a majority of scholars use Twitter, will of course generate different numbers of tweets compared with a more technologically conservative discipline.

*Future research*
The newness of altmetrics presents many opportunities for continued research. Most pressing is the need for environmental and content analysis surveys to build a fuller picture of how scholars use the tools and environment from which altmetrics draws. What does inclusion in a reference manger mean? Why and when do researchers bookmark content? How does linking to supporting papers from a blog post differ from citing in an editorial? These and similar questions need to be answered more fully before altmetrics can be used in any but the most cautious ways.

It will also be important to add more of a network perspective to altmetrics. Historically, bibliometricians have made surprisingly little use of network-analytic approaches. Things are changing in the citation literature, and this welcome trend should extend to altmetrics as well. Network-aware measurement will be particularly vital in informing recommendation and assessment algorithms. As Priego (2012) observes, the *number* of altmetrics events—tweets, for example—is rather less important than *who* they came from. Search engines faced similar problems in the early days of indexing the web; simply counting links was not sufficient, since some are more important than others. Approaches like Google's PageRank and related approaches, though, use the authority of linking pages—recursively calculated by the authority of other linking pages—to weight links. Similar approaches will be important for altmetrics; a tweet from a Nobel laureate is much more meaningful for many purposes than a tweet from the general public (as well as, arguably, more meaningful than a traditional citation).

Visualization holds promise for altmetrics as well; the multidimensionality and complexity of altmetrics makes their visualization more challenging, but also more important. The creation of composite metrics is another valuable research project. We might care less about the absolute numbers of, say, Mendeley and Facebook mentions, than we do about the ratio between them. Very low Facebook likes could be an indication of minimal popular interest that could make a high number of Mendeley bookmarks more compelling as a sign of scholarly value—a negative term in an academic impact regression equation. Similar relationships may exist for other different measures, or for more complex combinations of measures. Finally, altmetrics researchers should continue to expand the number and variety of data sources they investigate, examining scholarly impact from YouTube videos, Slideshare presentations, VIVO profiles, SSRN and Academia.edu downloads, Stack Overflow reputation, and more.

Altmetrics is a young, but growing, field. For now, there is perhaps more promise than results, but the promise is sufficient to justify further research. Those results we have seen give us good reason to believe that scholars, like other information workers, will continue to move their work into online tools and environments. To the extent that these tools allow us to peer into processes once hidden—and it seems this extent is significant—altmetrics will become increasingly important as a way to understand the hidden stories of scholarly impact, more directly observing the tracks of scholarly ideas.

Table 1. Altmetrics sources by type and audience

|  | General users | Scholarly users |
|---|---|---|
| Recommendation | web-based mainstream | Faculty of 1000 |

|  | media |  |
|---|---|---|
| Citation | Wikipedia | Citation from within peer-reviewed literature |
| Conversation | Twitter, Facebook, blogs | scholarly blogs, article comments, tweets from scholars |
| Reference | social bookmarking | social reference managers |
| Reading | html views | pdf downloads |

Figure 1: Growth in number of scholars actively using twitter, from (Priem et al., 2011) [I have a higher-resolution on I can email you, but it's too big for this doc]

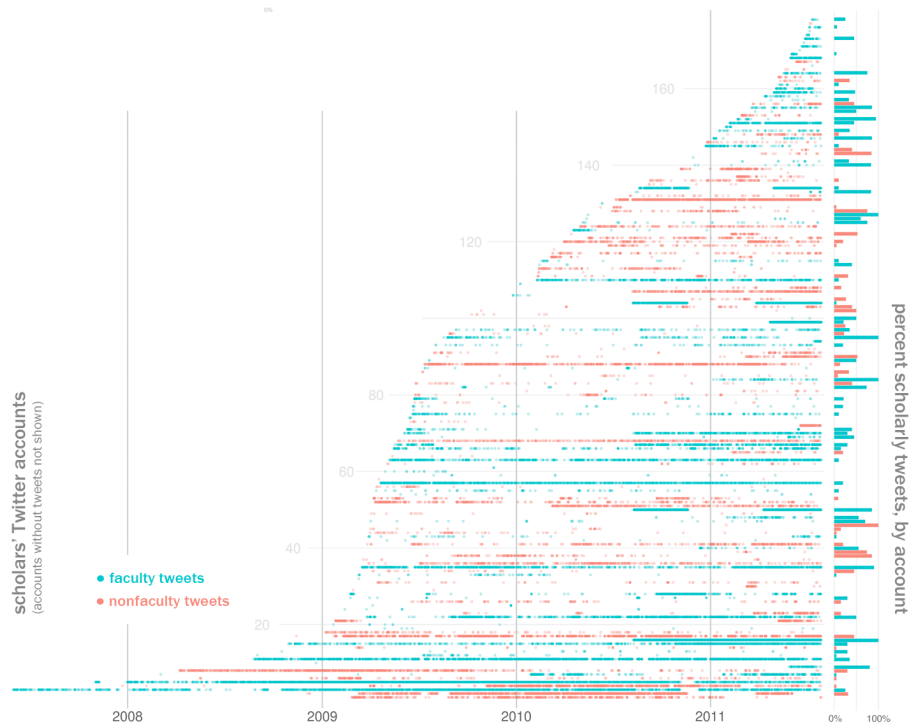

Figure 2: Article clusters based on altmetrics event types. Columns show the centers of clusters; rows represent altmetrics. Bluer cells indicate cluster centers on a relatively high standardized value for the given metric (Priem et al., 2012).

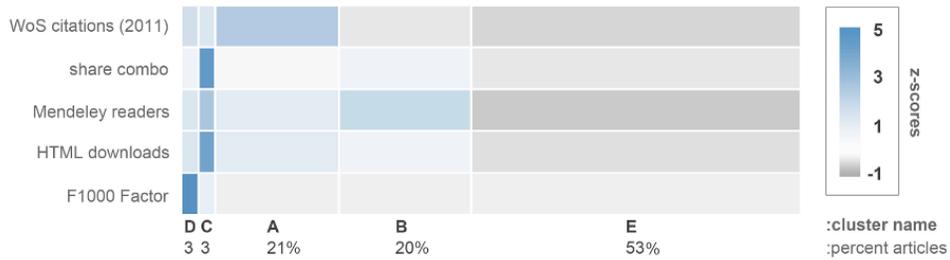

Figure 3: events relating to a single article, by time since the article's publication. Each dot represents one event. (Priem et al., 2012)

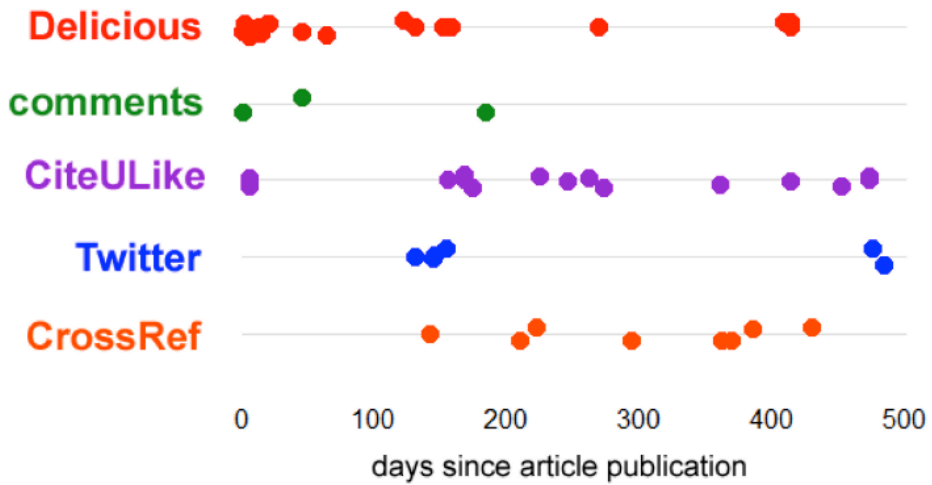